\def\al{ {\it et al}.~}
\def\rf#1{(\ref{#1})}
\def\lad{{\langle\hskip-3pt\langle}}
\def\rad{{\rangle\hskip-3pt\rangle}}
\def\la{\langle}
\def\ra{\rangle}
\def\lb{\left\{}
\def\rb{\right\}}
\def\Sk{{\bf S}^{(k)}}
\def\Smk{S_m^{(k)}}
\def\Gmk{{\bf\Gamma}^{(m,k)}}
\def\H#1{{\bf H}^{(#1)}}
\def\Q#1{{\bf Q}^{(#1)}}
\def\G#1{{\bf G}^{(#1)}}
\def\h#1{{\bf h}^{(#1)}}
\def\lmb#1{\lambda^{(#1)}}
\def\mi{\bigskip\noindent}
\documentstyle[12pt,psfig]{article}
\renewcommand{\thesection}{\arabic{section}.}

\begin{document}
{\center
{\it Short title: PLAN-FORM TWO-SCALE DYNAMOS}

\bigskip
{\bf CONVECTIVE PLAN-FORM TWO-SCALE DYNAMOS\\
IN A PLANE LAYER}

\mi V.A. Zheligovsky\footnote{E-mail: vlad@mitp.ru}

\mi International Institute of Earthquake Prediction Theory\\
and Mathematical Geophysics,\\
79 bldg.2, Warshavskoe ave., 117556 Moscow, Russian Federation

\mi Laboratory of general aerodynamics, Institute of Mechanics,\\
Lomonosov Moscow State University,\\
1, Michurinsky ave., 119899 Moscow, Russian Federation

\mi Observatoire de la C\^ote d'Azur, CNRS U.M.R. 6202,\\
BP 4229, 06304 Nice Cedex 4, France

\bigskip
Accepted in {\it Geophysical and Astrophysical Fluid Dynamics}

}

\bigskip{\bf Abstract.} We study generation of magnetic fields, involving large
spatial scales, by convective plan-forms in a horizontal layer.
Magnetic modes and their growth rates are expanded in power series
in the scale ratio, and the magnetic eddy diffusivity (MED) tensor is derived
for flows, symmetric about the vertical axis in a layer. For convective
rolls we demonstrate that MED is never below molecular magnetic diffusivity.
For cell patterns possessing the symmetries of a rectangle,
critical values of molecular magnetic
diffusivity for the onset of small- and large-scale magnetic field generation
are the same. No instances of negative MED in hexagonal cells have been detected.
A family of plan-forms has been found numerically, where MED is negative
for molecular magnetic diffusivity over the threshold for the onset of
small-scale magnetic field generation. However, the region in the parameter
space, where large-scale dynamo action is observed, is small.

\bigskip{\bf Key words.} Kinematic magnetic dynamo, magnetic modes,
symmetric flow, asymptotic expansion, magnetic eddy diffusivity.

\pagebreak
\section{Introduction}

The present work continues the series of studies of generation of
magnetic field possessing large spatial scales by Lanotte\al(2000),
Zheligovsky\al(2001) and Zheligovsky and Podvigina (2003).
Space-periodic parity-invariant flows, steady or periodic in time,
which have an exponentially decaying spectrum, were considered by these authors,
regarded as a model for natural turbulent flows of conducting fluids.
For such flows magnetic eddy diffusivity (MED) was found to be often negative.
(The analysis has been extended to the study of linear stability of
space periodic magnetohydrodynamic steady states to long-period perturbations
by Zheligovsky, 2003).

We consider here a kinematic dynamo problem in an infinite layer, assuming
perfectly conducting horizontal boundaries:\footnote{Roberts and Zhang (2000)
commented on physical validity of these boundary conditions.}
\begin{equation}
\left.{\partial H_1\over\partial x_3}\right|_{x_3=0,\pi}
=\left.{\partial H_2\over\partial x_3}\right|_{x_3=0,\pi}=0,
\quad H_3|_{x_3=0,\pi}=0.
\label{Hboundary}\end{equation}
As {\it ibid.}, the following algebraic idea serves as a foundation
for our constructions.

The kinematic dynamo problem for a steady flow ${\bf v}({\bf x})$ can be
reduced to an eigenvalue problem for the magnetic induction operator $\cal L$:
\begin{equation}
{\cal L}{\bf H}\equiv\eta\nabla^2{\bf H}+\nabla\times({\bf v}\times{\bf H})=\lambda{\bf H}.
\label{eigeneq}\end{equation}

The adjoint operator is
$${\cal L}^*{\bf H}\equiv\eta\nabla^2{\bf H}-{\bf v}\times(\nabla\times{\bf H}),$$
implying ${\cal L}^*{\bf H}=0$ for any constant vector field $\bf H$.
Hence, any constant vector field, satisfying the boundary conditions
for the adjoint operator, belongs to its kernel. (By the definition of
the adjoint operator, the identity
\begin{equation}
({\cal L}{\bf H}_1,{\bf H}_2)=({\bf H}_1,{\cal L}^*{\bf H}_2)
\label{defadj}\end{equation}
holds for all vector fields ${\bf H}_1({\bf x})$ and ${\bf H}_2({\bf x})$
from domains of $\cal L$ and ${\cal L}^*$, respectively; here the standard
scalar product $(\cdot,\cdot)$ of the functional Hilbert space $L^2$ (the
Lebesgue space) is assumed.
Boundary conditions for the adjoint operator can be determined demanding
that all surface integrals appearing in \rf{defadj} after integration by parts vanish.)
Generically the kernel of ${\cal L}^*$ is spanned by such constant vector fields.
Thus generically there exist as many neutral magnetic
modes (i.e. magnetic fields, satisfying \rf{eigeneq} with $\lambda=0$),
as there exist linearly independent constant vector fields satisfying
the boundary conditions for ${\cal L}^*$.

The flow is supposed to be periodic in horizontal directions:
\begin{equation}
{\bf v}({\bf x})={\bf v}\left(x_1+{2\pi\over L_1},x_2,x_3\right)=
{\bf v}\left(x_1,x_2+{2\pi\over L_2},x_3\right).
\label{vperiod}\end{equation}
If vector fields in the domain of $\cal L$
have the same periodicity \rf{vperiod} as the flow,
\begin{equation}
{\bf H}({\bf x})={\bf H}\left(x_1+{2\pi\over L_1},x_2,x_3\right)=
{\bf H}\left(x_1,x_2+{2\pi\over L_2},x_3\right),
\label{Hperiod}\end{equation}
(vector fields with such periodicity will be called {\it small-scale}) and satisfy
boundary conditions \rf{Hboundary}, then vector fields in the domain of
${\cal L}^*$ also satisfy \rf{Hperiod} and \rf{Hboundary}, provided
the fluid does not penetrate through horizontal boundaries of the layer:
\begin{equation}
v_3|_{x_3=0,\pi}=0.
\label{impenetr}\end{equation}
Hence in this case a constant vector field belongs to the kernel of
${\cal L}^*$, if and only if its vertical component vanishes. Thus,
$${\rm dim\ ker\ }{\cal L}^*\ge 2$$
and generically there exist two neutral magnetic modes
satisfying \rf{Hperiod} and \rf{Hboundary}.

The conducting fluid resides in an infinite volume.
When the spatial period of magnetic field is allowed to increase to infinity,
the smallest eigenvalue of the Laplacian tends to zero -- diffusive dissipation
of slowly varying magnetic fields is small. Therefore, the following question
is legitimate: Can a growing magnetic mode be constructed
by perturbing a neutral mode of the same periodicity as that of the flow,
allowing larger spatial scales in the perturbation?
It was found in the studies cited above, that growing modes can indeed be
constructed following this idea and using multiscale techniques.
We show in the present paper,
that this remains true for magnetic field generation in a layer.

The mechanism of negative MED for generation of large-scale magnetic field
requires that no $\alpha$-effect is present. To ensure this,
parity-invariant flows have been considered in the earlier studies.
In the present work a different symmetry is employed for this purpose --
the symmetry about the vertical axis:
$$v_1(-x_1,-x_2,x_3)=-v_1(x_1,x_2,x_3),$$
\begin{equation}
v_2(-x_1,-x_2,x_3)=-v_2(x_1,x_2,x_3),
\label{symmaxis}\end{equation}
$$v_3(-x_1,-x_2,x_3)=v_3(x_1,x_2,x_3).$$
This type of symmetry is not unnatural: it is sustained by the Navier-Stokes
equation (if the external forces involved have this symmetry).

In Section 2 we discuss expansions of magnetic modes and their growth rates
in power series in scale ratio, $\epsilon$, which is a small parameter
of the problem, and present MED tensor for a general incompressible flow
in a layer satisfying \rf{vperiod} and \rf{impenetr}.
A complete formal expansion of magnetic modes and the associated eigenvalues
is exposed in detail in Appendix A. In Section 3 we prove that for plane
flows (and hence for convective rolls) eddy correction to molecular magnetic
diffusivity is always non-negative. In the remaining part of the paper results
of Section 2 are applied to convective plan-forms without rotation.
Relevance of the kinematic dynamo
problem for convective plan-forms for the study of the full non-linear
magnetohydrodynamic system has been discussed in the context of
the theory of bifurcations by Bosh-Vivancos\footnote{Numerical results
of this paper have been questioned by Zheligovsky and Galloway
(1998) and Matthews (1999b).}\al(1995).
Representation and symmetry properties of convective plan-forms are summarised
in Section 4. The plan-forms possess additional symmetries, for which the MED
tensor is diagonal, as discussed in Section 5. In Section 6 numerical
results for two families of plan-forms are presented.
For cell patterns possessing the symmetries of a rectangle,
the region in the parameter space of negative MED
coincides with the region of small-scale magnetic field generation.
A family of cells with a smaller symmetry group
has been found numerically, where MED is negative
for molecular magnetic diffusivity over the threshold for the onset of
small-scale magnetic field generation. In Appendix B
we consider alternative representations of elements of the MED tensor
and procedures for computation of MED, which are more efficient
than straightforward numerical solution of
auxiliary problems arising in construction of the asymptotic series.

\section{Magnetic eddy diffusivity tensor\hfill\break
for large-scale magnetic modes in a layer}

We outline here construction of expansions of magnetic modes and their growth
rates. This is done under {\it the basic assumption} that a generic problem
is considered, i.e. the kernel of the adjoint operator ${\cal L}^*$
is spanned by constant vector fields, whose vertical component vanishes.
In this Section a steady fluid flow ${\bf v}({\bf x})$ is supposed\\
$i$. to satisfy the boundary conditions \rf{vperiod} and \rf{impenetr};\\
$ii$. to possess the symmetry \rf{symmaxis};\\
$iii$. to have a zero space average of the horizontal components;\\
$iv$. to be solenoidal:
\begin{equation}
\nabla\cdot{\bf v}=0.
\label{vsolenoid}\end{equation}

A magnetic mode ${\bf H}({\bf x},{\bf y})$ is supposed to depend on the
{\it fast} variable ${\bf x}\in R^3$ and on the {\it slow} variable
${\bf y}=\epsilon(x_1,x_2)$ in horizontal directions.
Here the scale ratio $\epsilon>0$ is a small parameter.
A solution to the eigenvalue problem \rf{eigeneq} is sought in the form
of power series in $\epsilon$:
\begin{equation}
{\bf H}=\sum_{n=0}^\infty\h{n}({\bf x},{\bf y})\epsilon^n,
\label{Hseries}\end{equation}
\begin{equation}
\lambda=\sum_{n=0}^\infty\lmb{n}\epsilon^n.
\label{lambdaseries}\end{equation}
The mode is supposed to be $2\pi/L_j$-periodic in $x_j$ for $j=1,2$,
and to satisfy \rf{Hboundary} on the horizontal boundaries $x_3=0$ and
$x_3=\pi$. Accordingly, we assume that each $\h{n}$ satisfies \rf{Hperiod}
and \rf{Hboundary}.

Let $\la\cdot\ra$ and $\lad\cdot\rad$ denote the mean part of a scalar or
vector field and the mean horizontal part of a vector field, respectively:
$$\la f\ra\equiv{1\over V}\int_0^{\pi}\int_{-{\pi/L_2}}^{\pi/L_2}
\int_{-{\pi/L_1}}^{\pi/L_1}f({\bf x},{\bf y}){\rm d}{\bf x},\qquad
\lad{\bf f}\rad\equiv\sum_{j=1}^2\la{\bf f}\cdot{\bf e}_j\ra{\bf e}_j.$$
Here $V=4\pi^3/(L_1L_2)$ is the volume of the periodicity box, and ${\bf e}_j$
is a unit vector in the direction along the coordinate axis $x_j$. Denote
$$\H{n}=\lad\h{n}\rad.$$

By the chain rule, dependence of the magnetic mode on the fast and slow
variables implies that gradients must be modified in the eigenvalue equation
\rf{eigeneq}:
\begin{equation}
\nabla\to\nabla_{\bf x}+\epsilon\nabla_{\bf y},\quad{\rm where }\quad
\nabla_{\bf y}\equiv\left[{\partial\over\partial y_1},{\partial\over\partial y_2},0\right].
\label{modnabla}\end{equation}
Substituting \rf{Hseries}-\rf{modnabla} into \rf{eigeneq}, one obtains a
hierarchy of equations, which can be solved successively in all orders
of $\epsilon$ (see Appendix A).

In particular, one finds $\lambda=O(\epsilon^2):\ \lmb{0}=\lmb{1}=0$, and
$\lmb{2}$ is determined from the eigenvalue equation for the mean horizontal
part of the leading term in the expansion of the magnetic mode \rf{Hseries}:
\begin{equation}
{\cal M}\H{0}\equiv\eta\nabla^2_{\bf y}\H{0}+{\cal P}\nabla_{\bf y}\times
\sum_{k=1}^2\sum_{m=1}^2\la{\bf v}\times\Gmk\ra
{\partial H_k^{(0)}\over\partial y_m}=\lmb{2}\H{0}.
\label{eddyeigen}\end{equation}
Here $\cal P$ is the projector onto horizontal directions:
$${\cal P}{\bf f}=\sum_{j=1}^2({\bf f}\cdot{\bf e}_j){\bf e}_j.$$
$\cal M$ is called the {\it operator of (anisotropic) MED}.
Its coefficients can be determined from two auxiliary problems:\\
{\it the first auxiliary problem:}
\begin{equation}
{\cal L}{\bf S}^{(k)}=-{\partial{\bf v}\over\partial x_k}\quad(k=1,2);
\label{Sproblem}\end{equation}
{\it the second auxiliary problem:}
\begin{equation}
{\cal L}\Gmk=-2\eta{\partial\Sk\over\partial x_m}
-{\bf e}_m\times({\bf v}\times(\Sk+{\bf e}_k))\quad(m,k=1,2).
\label{Gproblem}\end{equation}
Vector fields $\Sk$ and $\Gmk$ satisfy the boundary conditions
\rf{Hperiod} and \rf{Hboundary}. The basic assumption (stated in the beginning
of this Section) implies solvability of the problems \rf{Sproblem}
and \rf{Gproblem}. It is shown in Appendix A that
\begin{equation}
\nabla\cdot\Sk=0,
\label{Ssolenoid}\end{equation}
\begin{equation}
\nabla\cdot\Gmk+S^{(k)}_m=0.
\label{Gsolenoid}\end{equation}
Alternative expressions for elements of the MED tensor can be obtained,
partially performing integration required for averaging of the cross-products
of $\bf v$ and $\Gmk$. They are derived in Appendix B.

$\cal M$ is a second order operator in partial derivatives with constant
coefficients. Consequently, solutions to \rf{eddyeigen} bounded on the
entire plane are Fourier harmonics
\begin{equation}
\H{0}={\bf h}{\rm e}^{{\rm i}\bf q\cdot y},
\label{Fourierharm}\end{equation}
where ${\bf q}\in R^2$ is an arbitrary constant wave vector,
and ${\bf h}\in R^3$ is a constant vector satisfying
\begin{equation}
\eta|{\bf q}|^2{\bf h}+{\cal P}\left([q_1,q_2,0]\times
\sum_{k=1}^2\sum_{m=1}^2\la{\bf v}\times\Gmk\ra q_mh_k\right)=-\lmb{2}\bf h
\label{toyeigen}\end{equation}
and
\begin{equation}
h_3=0.
\label{toyflat}\end{equation}
The solenoidality condition for the magnetic mode
\begin{equation}
\nabla\cdot{\bf H}=0
\label{Hsolenoid}\end{equation}
implies
$\nabla_{\bf y}\cdot\H{0}=0$ (see Appendix A), hence
\begin{equation}
[q_1,q_2,0]\cdot{\bf h}=0.
\label{toysolenoid}\end{equation}
From \rf{toyeigen}, \rf{toyflat} and \rf{toysolenoid},
\begin{equation}
{\bf h}=[-q_2,q_1,0],\qquad
-\lmb{2}({\bf q})=\eta|{\bf q}|^2-\sum_{k=1}^2\sum_{m=1}^2
(-1)^k\la{\bf v}\times\Gmk\ra_3\,q_mq_{3-k}.
\label{toysolved}\end{equation}

If $|{\bf q}|=1$, following the long established tradition $-\lmb{2}$
is called {\it magnetic eddy diffusivity}. If the minimal MED is negative:
$$\min_{|{\bf q}|=1}(-\lmb{2}({\bf q}))<0,$$
there exist
growing large-scale magnetic modes, i.e. the flow operates as a dynamo.

The leading term of the magnetic mode expansion \rf{Hseries} can be expressed as
$$\h{0}({\bf x},{\bf y})={\rm e}^{{\rm i}\bf q\cdot y}\sum_{k=1}^2h_k(\Sk({\bf x})+{\bf e}_k).$$
Since \rf{Sproblem} is equivalent to
\begin{equation}
{\cal L}(\Sk+{\bf e}_k)=0,
\label{Skernel}\end{equation}
the magnetic mode is a perturbation of the neutral small-scale magnetic mode
$$\sum_{k=1}^2h_k(\Sk({\bf x})+{\bf e}_k)\in\ker{\cal L}$$
modulated by an amplitude factor depending on the slow variable.

\section{Magnetic eddy diffusivity for plane parallel flows}

By the Zeldovich (1956) theorem (see also Moffatt, 1978), plane flows (such
flows $\bf v(x)$ that ${\bf v(x)\cdot k}=0$ for a constant vector $\bf k$)
cannot generate magnetic field. Here a stronger in some sense result is derived:
plane parallel flows can only enhance molecular diffusivity. More precisely,
we show that magnetic eddy correction for plane parallel flows in a layer,
which satisfy conditions $i-iv$ stated in the first paragraph of Section 2,
is always non-negative. Without any loss of generality we assume
in this Section, that the flow does not depend on $x_2$ and $v_2=0$.

\underline{$k=1$.} From \rf{Sproblem},
$$S^{(1)}_2=0\quad\Rightarrow\quad
{\bf v}\times({\bf S}^{(1)}+{\bf e}_1)=R({\bf x}){\bf e}_2.$$
Thus for $m=1$ the right-hand side of \rf{Gproblem} equals
$-2\eta\,\partial{\bf S}^{(1)}/\partial x_1-R({\bf x}){\bf e}_3$;
consequently
$$\Gamma^{(1,1)}_2=0\quad\Rightarrow\quad
\la{\bf v}\times{\bf\Gamma}^{(1,1)}\ra_3=0.$$
For $m=2$, the right-hand side of \rf{Gproblem} vanishes and hence
$${\bf\Gamma}^{(2,1)}=0\quad\Rightarrow\quad\la{\bf v}\times{\bf\Gamma}^{(2,1)}\ra_3=0.$$

\underline{$k=2$.} From \rf{Sproblem}, ${\bf S}^{(2)}=0$.
Hence, for $m=1$ the right-hand side of \rf{Gproblem} is $v_1{\bf e}_2$.
This implies ${\bf\Gamma}^{(1,2)}=\Gamma{\bf e}_2$,
where $\Gamma$ satisfies
$$\eta\nabla^2\Gamma-({\bf v}\cdot\nabla)\Gamma=v_1.$$
A similar equation arises in the multiscale
analysis of the passive scalar transport equation (see Biferale\al, 1995).
The respective coefficient of the MED tensor is non-negative:
$$-\la{\bf v}\times{\bf\Gamma}^{(1,2)}\ra_3=
-\la v_1\Gamma\ra=\eta\la|\nabla\Gamma|^2\ra\ge0$$
(it is positive for ${\bf v}\ne0$).

For $m=2$, the right-hand side of \rf{Gproblem} is equal to $-\bf v$, and hence
$${\bf\Gamma}^{(2,2)}_2=0\quad\Rightarrow\quad
\la{\bf v}\times{\bf\Gamma}^{(2,2)}\ra_3=0.$$

Consequently, \rf{toysolved} implies
$$-\lmb{2}=\eta|{\bf q}|^2-\la{\bf v}\times{\bf\Gamma}^{(1,2)}\ra_3\,q_1^2
\ge\eta|{\bf q}|^2,$$
and thus minimal MED is equal to molecular magnetic diffusivity $\eta$,
as we intended to demonstrate.

\section{Convective plan-form flows in a layer}

Analysis of Section 2 will be applied to convective plan-forms without rotation.
In this Section we summarise some of their properties (see Chandrasekhar, 1981).
Thermal convection in a fluid heated from below in a layer with no rotation
is considered. Plan-forms are instability modes (more precisely, the
flow parts of the modes) at the onset of instability of the trivial steady state
(in which the fluid is at rest and the temperature profile is linear in $x_3$).
They are poloidal:
\begin{equation}
{\bf v}({\bf x})\equiv\nabla\times\nabla\times(P({\bf x}){\bf e}_3).
\label{planforms}\end{equation}
In the most general form (Bisshopp, 1960) the potential can be expressed as
\begin{equation}
P({\bf x})\equiv(\alpha_1\cos(L_1x_1)\cos(L_2x_2)+\alpha_2\cos(pL_2x_2))w(x_3).
\label{potential}\end{equation}
Here $\alpha_1$ and $\alpha_2$ are constant and $p$ is integer.
For $\alpha_2\ne0$,
$$L_1/L_2=\sqrt{p^2-1}.$$

Evidently, plan-forms satisfy conditions $ii-iv$ assumed in Section 2.
Fluid does not penetrate through the horizontal boundaries (condition $i$),
if and only if
\begin{equation}
w(x_3)|_{x_3=0,\pi}=0.
\label{planbound}\end{equation}
If in addition
$$\left.{\partial^{2m}w\over\partial x^{2m}_3}\right|_{x_3=0,\pi}=0\quad\forall m>0,$$
the plan-form satisfies the {\it free boundary} conditions:
$$\left.{\partial v_1\over\partial x_3}\right|_{x_3=0,\pi}
=\left.{\partial v_2\over\partial x_3}\right|_{x_3=0,\pi}=0,\quad
v_3|_{x_3=0,\pi}=0.$$
Consequently, for free boundaries the vertical profile is
\begin{equation}
w_n(x_3)=\sin nx_3,
\label{freeplan}\end{equation}
where $n>0$ is integer. The mode for $n=1$ is the first to become unstable
when the Rayleigh number is increased. For rigid horizontal boundaries
the vertical profile is more complex (see Chandrasekhar, 1981).

Plan-forms possess two ``reflection" symmetries, referred to in
the next Section:\\
reflection in the direction ${\bf e}_1$:
$$v_1(-x_1,x_2,x_3)=-v_1(x_1,x_2,x_3),$$
\begin{equation}
v_2(-x_1,x_2,x_3)=v_2(x_1,x_2,x_3),
\label{xreflection}\end{equation}
$$v_3(-x_1,x_2,x_3)=v_3(x_1,x_2,x_3);$$
reflection in the direction ${\bf e}_2$:
$$v_1(x_1,-x_2,x_3)=v_1(x_1,x_2,x_3),$$
\begin{equation}
v_2(x_1,-x_2,x_3)=-v_2(x_1,x_2,x_3),
\label{yreflection}\end{equation}
$$v_3(x_1,-x_2,x_3)=v_3(x_1,x_2,x_3).$$
(The symmetry about the vertical axis is a composition of these two.)

Free boundary plan-forms with the vertical profile \rf{freeplan} are
parity-invariant about centers specified in the following table:

\bigskip
\renewcommand{\arraystretch}{2.4}
\hfill~\begin{tabular}{|c|c|c|}\hline
$n$&$p$&Centers\\\hline
odd & $\alpha_2=0$ &$\displaystyle\left({\pi\over L_1}
\left(l_1+{1\over2}\right),{\pi l_2\over L_2},{\pi\over2}\right)$\\
odd & odd, or $\alpha_2=0$ &$\displaystyle\left({\pi l_1\over L_1},
{\pi\over L_2}\left(l_2+{1\over2}\right),{\pi\over2}\right)$\\
even & any &$\displaystyle
\left({\pi l_1\over L_1},{\pi l_2\over L_2},{\pi\over2}\right)$\\
even & even, or $\alpha_2=0$ &$\displaystyle
\left({\pi\over L_1}\left(l_1+{1\over2}\right),
{\pi\over L_2}\left(l_2+{1\over2}\right),{\pi\over2}\right)$\\
\hline\end{tabular}\hfill~
\renewcommand{\arraystretch}{1.}

\mi
Here $l_1$ and $l_2$ are arbitrary integers. For such plan-forms,
the same asymptotic expansions could be constructed using parity
invariance instead of \rf{symmaxis}.

\section{The operator of magnetic eddy diffusivity\hfill\break
for convective plan-form flows in a layer}

Each of the two reflection symmetries
\rf{xreflection} and \rf{yreflection} splits the domain of $\cal L$ into
a direct sum of two proper subspaces: symmetric or antisymmetric with respect to the
given symmetry. (A vector field is said to possess the reflection antisymmetry
in the direction ${\bf e}_1$, if it satisfies \rf{xreflection} with the
reversed signs in the right-hand side:
$$v_1(-x_1,x_2,x_3)=v_1(x_1,x_2,x_3),$$
$$v_2(-x_1,x_2,x_3)=-v_2(x_1,x_2,x_3),$$
$$v_3(-x_1,x_2,x_3)=-v_3(x_1,x_2,x_3);$$
a reflection antisymmetry in the direction ${\bf e}_2$ is defined similarly.)

Consequently, due to \rf{Sproblem} the solution $\Sk$ to the first
auxiliary problem has the same reflection symmetries/antisymmetries,
as the vector field $\displaystyle\partial{\bf v}/\partial x_k$.
Furthermore, due to \rf{Gproblem} the solution $\Gmk$ to the second
auxiliary problem has the same reflection symmetries/antisymmetries,
as $\displaystyle\partial^2{\bf v}/\partial x_m\partial x_k$.

In particular, ${\bf\Gamma}^{(k,k)}$ possess both reflection symmetries and hence
$$\la{\bf v}\times{\bf\Gamma}^{(k,k)}\ra_3=0.$$
This implies diagonality of the MED operator,
restricted to the subspace of solenoidal vector fields
of slow variables with the zero vertical component:
$${\cal M}=(\eta-\la{\bf v}\times\Gamma^{(1,2)}\ra_3){\partial^2\over\partial y_1^2}
+(\eta+\la{\bf v}\times\Gamma^{(2,1)}\ra_3){\partial^2\over\partial y_2^2}.$$
Therefore, for a flow possessing the symmetries \rf{xreflection} and
\rf{yreflection}, the minimal MED is
\begin{equation}
\min_{|{\bf q}|=1}(-\lmb{2}({\bf q}))=
\min(\eta-\la{\bf v}\times{\bf\Gamma}^{(1,2)}\ra_3,
\eta+\la{\bf v}\times{\bf\Gamma}^{(2,1)}\ra_3).
\label{mineddy}\end{equation}

\section{Numerical results}

As shown in the previous Section, to find coefficients of the MED tensor
and thus the minimal MED for a plan-form it appears necessary to solve
four problems ${\cal L}{\bf X}=\bf F$: two first, and two second auxiliary
problems \rf{Sproblem} and \rf{Gproblem}. However, as discussed in Appendix B,
it suffices to solve three elliptic problems of the same complexity.

Computations have been performed for flows \rf{planforms}, \rf{potential}
with the vertical profile
\begin{equation}
w(x_3)=\sum_{n=1}^N\beta_n\sin(n x_3),
\label{myw}\end{equation}
where $\beta_n$ are constant coefficients. The flows employed in computations
are normalised, so that the root mean square is equal to 1.
All Figures below show results obtained for plan-forms
with the vertical profile \rf{freeplan}.

\subsection{Fourier representation of the vector fields $\Sk$ and
$\Gmk$}

Since solutions to the auxiliary problems possess the reflection symmetries
\rf{xreflection} and \rf{yreflection} (see Section 5) and satisfy the boundary
conditions \rf{Hboundary}, they can be expanded in the following Fourier series:
\begin{equation}
{\bf S}^{(1)}=\sum_{n_i\ge0}\left[
\begin{array}{c}
s^{(1)}_{{\bf n},1}\cos n_1L_1x_1\ \cos n_2L_2x_2\ \cos n_3x_3\\
s^{(1)}_{{\bf n},2}\sin n_1L_1x_1\ \sin n_2L_2x_2\ \cos n_3x_3\\
s^{(1)}_{{\bf n},3}\sin n_1L_1x_1\ \cos n_2L_2x_2\ \sin n_3x_3
\end{array}\right],
\label{S1sum}\end{equation}
\begin{equation}
{\bf S}^{(2)}=\sum_{n_i\ge0}\left[
\begin{array}{c}
s^{(2)}_{{\bf n},1}\sin n_1L_1x_1\ \sin n_2L_2x_2\ \cos n_3x_3\\
s^{(2)}_{{\bf n},2}\cos n_1L_1x_1\ \cos n_2L_2x_2\ \cos n_3x_3\\
s^{(2)}_{{\bf n},3}\cos n_1L_1x_1\ \sin n_2L_2x_2\ \sin n_3x_3
\end{array}\right],
\label{S2sum}\end{equation}
\begin{equation}
\Gmk=\sum_{n_i\ge0}\left[
\begin{array}{c}
\gamma^{(m,k)}_{{\bf n},1}\cos n_1L_1x_1\ \sin n_2L_2x_2\ \cos n_3x_3\\
\gamma^{(m,k)}_{{\bf n},2}\sin n_1L_1x_1\ \cos n_2L_2x_2\ \cos n_3x_3\\
\gamma^{(m,k)}_{{\bf n},3}\sin n_1L_1x_1\ \sin n_2L_2x_2\ \sin n_3x_3
\end{array}\right].
\label{Gammasum}\end{equation}

All computations have been performed with the resolution of 64 trigonometric
functions in each direction. With this resolution the energy spectrum of
solutions decays by several (at least 4) orders of magnitude.

The following parity symmetries each reduce twice the number of
unknown coefficients. If in \rf{potential} $p$ is odd or $\alpha_2=0$,
the sum of horizontal wave numbers in the velocity is even; hence vector fields
constructed of harmonics, where the sum of the wave numbers in the horizontal
directions is either even or odd, constitute, respectively, two invariant subspaces
of the domain of $\cal L$. Therefore, if $p$ is odd or $\alpha_2=0$,
coefficients of the series \rf{S1sum}-\rf{Gammasum} for odd $n_1+n_2$ vanish.
Similarly, if in \rf{myw} only odd wave-number terms are present, coefficients
of the series \rf{S1sum}-\rf{Gammasum} for odd $n_1+n_3$ vanish.

For plan-forms with the vertical profile \rf{freeplan} for $n>1$,
coefficients of the series \rf{S1sum}-\rf{Gammasum} vanish for all $n_3$,
which are not divisible by $n$, and the second parity symmetry is modified:
coefficients are non-zero only for even $n_1+n_3/n$.

\centerline{\psfig{file=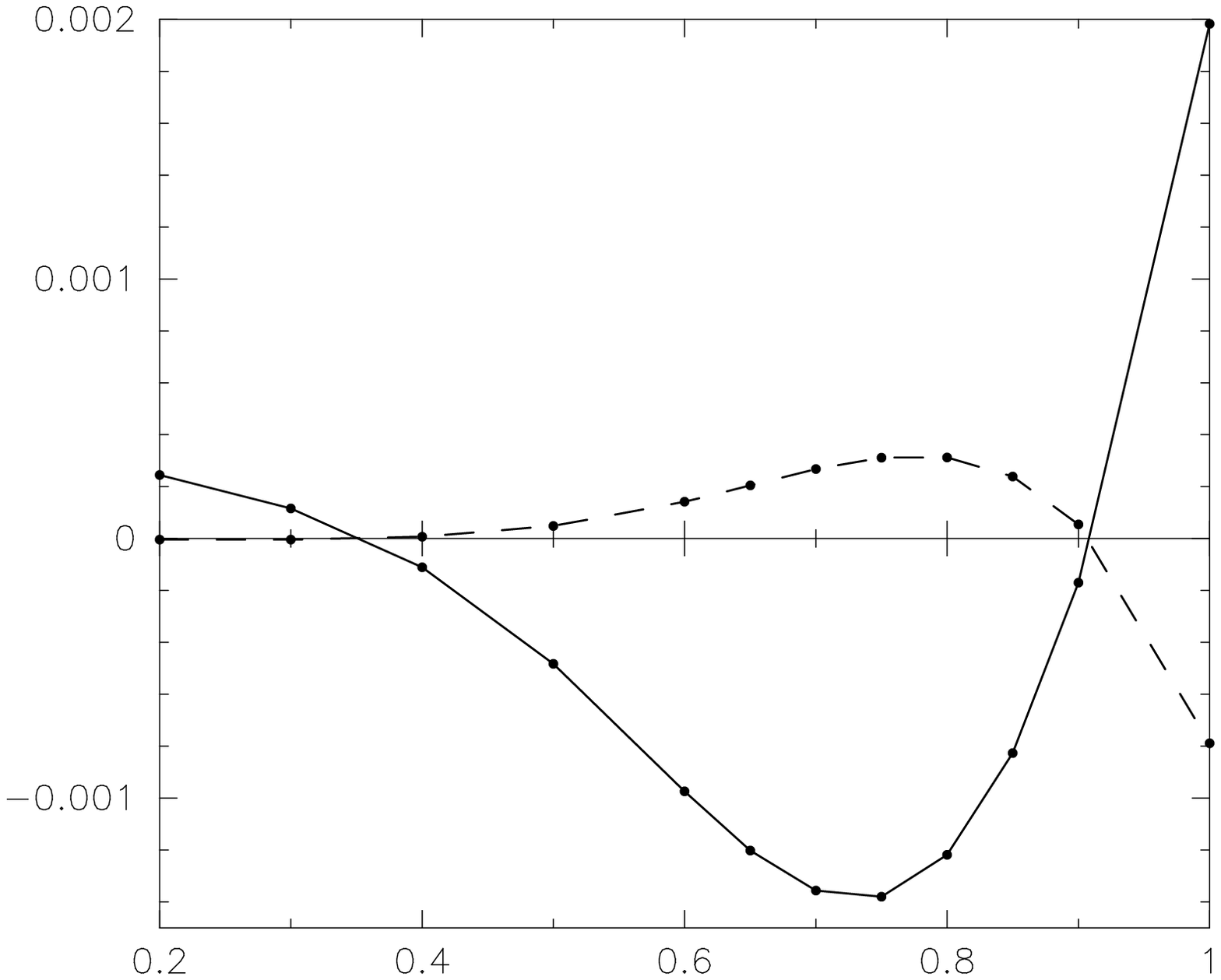,width=14cm,clip=}}

\mi
Figure 1. Minimal MED (solid line; vertical axis) and the growth rate of
the dominant small-scale magnetic mode with a zero horizontal mean
(dashed line; vertical axis) for cell patterns possessing the symmetries of
a rectangle, for $L_2=2,\ \eta=0.06$, as a function of $L_1$
(horizontal axis). Computed values are shown by solid dots.

\subsection{MED for rectangular cell patterns}

Thermal convection cell patterns possessing the symmetries of a rectangle
are plan-forms \rf{planforms}, \rf{potential}, where $\alpha_2=0$.

Matthews (1999a,b)\footnote{Opposite to Matthews (1999b), we cannot recommend
integration of the magnetic induction equation in time as an {\it efficient}
technique for numerical determination of dominant magnetic modes
for steady flows, at least unless the following improvements are implemented:
1) Integration of the magnetic induction equation in time is performed
using fast specialised time-stepping schemes overcoming
stiffness of the problem (e.g., see Nikitin 1994, 1996).
(They can be easily implemented, if the Galerkin discretisation of the induction
equation in space in the basis of eigenfunctions of the Laplacian is employed.)
2) Optimisation methods (e.g., similar to those of Zheligovsky, 1993)
are employed, allowing one to jump from the current trajectory of temporal
evolution of a magnetic field to a different one, closer to the exponential
trajectory for the dominant magnetic mode.
3) Smallness of the discrepancy $|{\cal L}{\bf h}-\xi{\bf h}|$, and not
the ``overall exponential behaviour" of the obtained solution
is examined in the condition for termination of computations.

In the present work a spectral code derived from that of Zheligovsky (1993)
has been used.}
found square plan-forms with $L_1=L_2=1/2$ incapable of kinematic
dynamo action. He assumed the boundary conditions \rf{Hboundary} and
considered magnetic modes with vectors of basic periods along the diagonals
of our basic periodicity square, with the periods equal to a half of the
diagonal length (constituting a subspace of our small-scale magnetic fields).

\begin{figure}
\centerline{\psfig{file=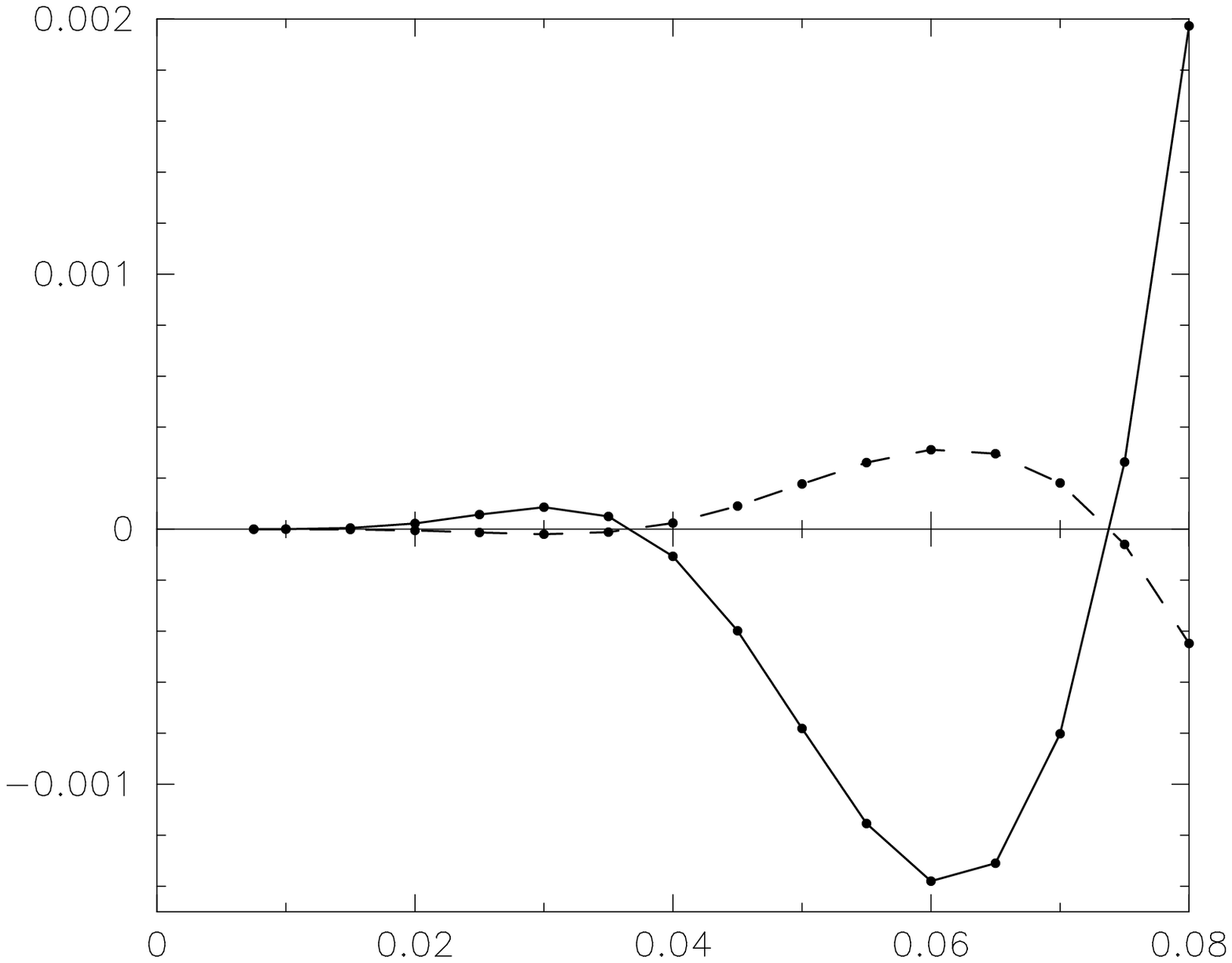,width=14cm,clip=}}

\mi
Figure 2. Minimal MED (solid line; vertical axis) and the growth rate of
the dominant small-scale magnetic mode with a zero horizontal mean
(dashed line; vertical axis) for cell patterns possessing the symmetries of
a rectangle, for $L_1=0.75,\ L_2=2$, as a function of $\eta$
(horizontal axis). Computed values are shown by solid dots.
\end{figure}

Figures 1 and 2 display MED computed for two sets of parameter values
for plan-forms  with the symmetry group of a rectangle,
together with the growth rate of the dominant
magnetic mode of the same spatial periodicity as in the flow
(the associated eigenvalues of $\cal L$ are real).

The Figures illustrate an unexpected phenomenon:
The critical molecular viscosity for the onset of small-scale magnetic field
generation coincides with that for the onset of large-scale magnetic field
generation; i.e. the minimal MED vanishes together with the growth rate
of the dominant magnetic mode of the spatial periodicity of the flow.
Since the growth rate of the large-scale magnetic mode is small (${\rm O}(\epsilon^2)$),
the large-scale magnetic instability is weak compared to
the small-scale one.

This phenomenon has been reproduced in simulations for a number of
vertical profiles \rf{myw} comprised of up to three sines with wave numbers
of the same or different parity. In these simulations the critical molecular
diffusivities $\eta$ have been determined to the accuracy of at least $10^{-8}$.
Algebraic reasons of this phenomenon are not entirely clear.

\begin{figure}
\centerline{\psfig{file=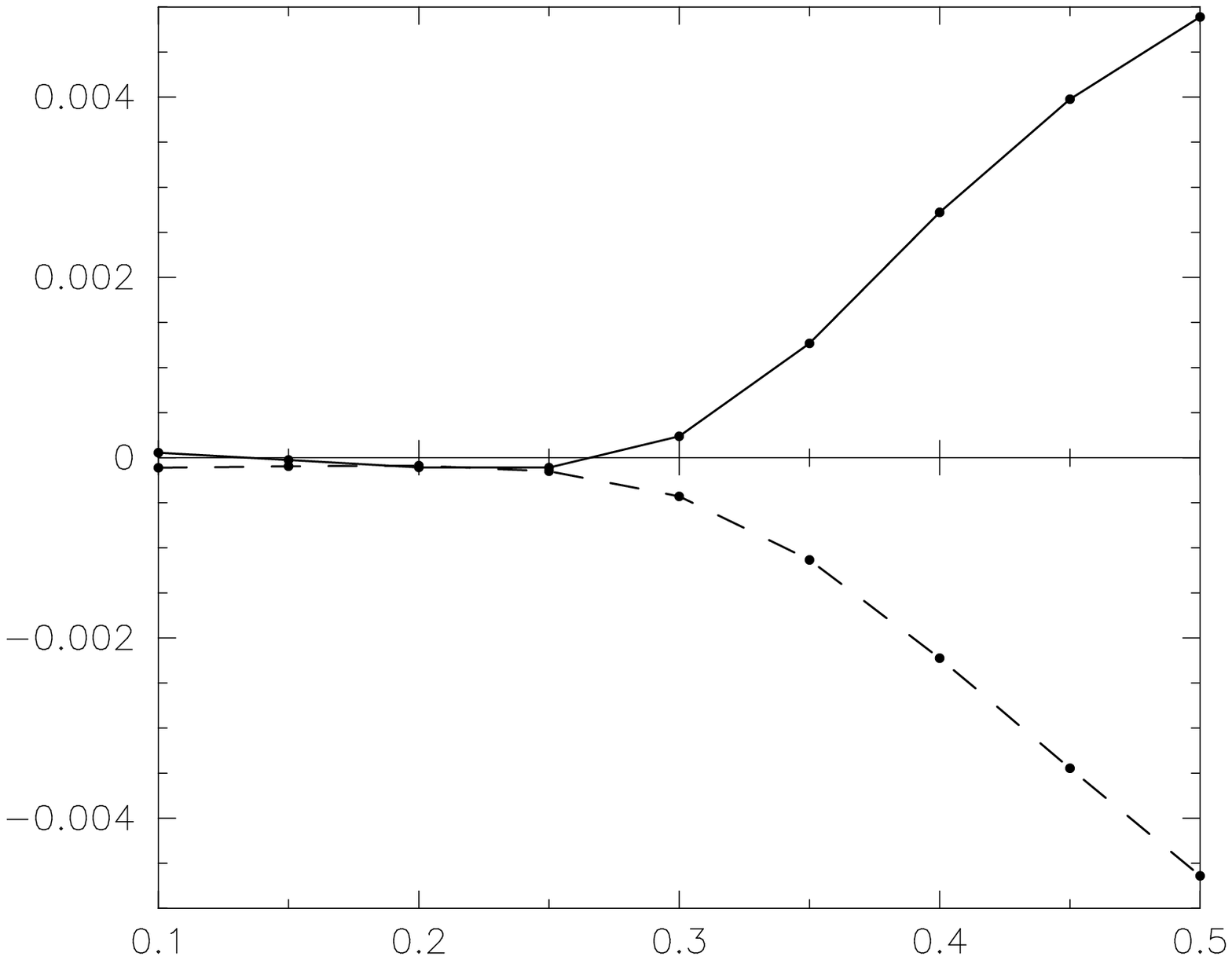,width=14cm,clip=}}

\mi
Figure 3. Minimal MED (solid line; vertical axis) and the growth rate of
the dominant small-scale magnetic mode with a zero horizontal mean
(dashed line; vertical axis) for plan-forms for
$p=2,\ L_1=1.5\sqrt{3},\ L_2=1.5,\ \eta=0.0075$ as a function of
$\alpha_2/\alpha_1$ (horizontal axis). Computed values are shown by solid dots.
\end{figure}

Suppose horizontal coordinate axes are such that
$\eta-\la{\bf v}\times{\bf\Gamma}^{(1,2)}\ra_3$ is the minimal MED
(cf.~\rf{mineddy}); they can be always chosen this way, since for $\alpha_2=0$
the two horizontal coordinate directions are interchangeable.
Then the neutral small-scale magnetic mode has the same reflection symmetries
as ${\bf\Gamma}^{(1,2)}$, and can be expanded in a Fourier series of the form of
\rf{Gammasum}. All non-vanishing terms in this series correspond to wave
vectors, whose $\bf n$ components have the parity either (odd, even, even), or
(even, odd, odd). This is in contrast with the structure of ${\bf\Gamma}^{(1,2)}$,
which is represented by the series \rf{Gammasum} where all non-vanishing terms
correspond to wave vectors, all three components of which have the same parity.

\subsection{MED for general cell patterns}

Figure 3 displays MED computed for a family of cell patterns
for $\alpha_2\ne 0$ (which do not possess therefore the symmetries
of a rectangle) together with the growth rate of the dominant magnetic mode of
the same spatial periodicity as in the flow (the associated eigenvalues
of $\cal L$ are again real). A window of small ratios $\alpha_2/\alpha_1$
has been detected, where MED is
negative. For the assumed value of molecular viscosity flows from this family
do not generate small-scale magnetic field; thus in this family
of plan-forms the effect of negative MED instability is separated out.

For $\alpha_2=\alpha_1/2$ the flow is the Christopherson (1940) hexagonal cell
pattern. Zheligovsky and Galloway (1998) found that in case there is a dielectric
beyond one of the horizontal boundaries of the layer, the flow can generate
small-scale magnetic field for small molecular diffusivities.
Matthews (1999b) has not found magnetic field generation by hexagonal cells
for $L_1=\sqrt{3/8},\ L_2=\sqrt{1/8}$, assuming the boundary conditions
\rf{Hboundary} and \rf{Hperiod}. A number of computations for the hexagonal
cell pattern have been performed for two-term sums \rf{myw} and a varying ratio
$\beta_n/\beta_1$, but no instances of negative MED have been found.

\section{Conclusion}
Formal asymptotic expansions of magnetic modes and the associated eigenvalues
in series in the scale ratio have been constructed for flows, symmetric about
the vertical axis. The region where the flows reside and their symmetries are
different from those for parity-invariant space-periodic flows, which have
been considered so far. If a flow possesses reflection symmetries about
the horizontal Cartesian axes, the magnetic eddy diffusivity tensor becomes
diagonal. The suggested strategies for evaluation of elements of the MED
tensor require numerical solution of a smaller number of elliptic PDE's,
than for the direct approach whereby all auxiliary problems are solved.
Numerical experiments have been performed for convective plan-forms in the
absence of rotation, and a family of plan-forms exhibiting negative MED has
been found; these flows cannot generate small-scale magnetic field with
the spatial periodicity of the flow. Therefore magnetic field generation
in a layer by the mechanism of negative MED is possible. The flows
are poloidal, and their helicity vanishes identically. Thus, the paradigm
requiring helicity of a flow to be non-zero to enable it to act as a dynamo
has been once again demonstrated to be wrong. However, the region
in the parameter space, where large-scale dynamo operates, is small, and
the minimal negative MED's found in computations are small in absolute value.

\section{\bf Acknowledgments}

I am indebted to Prof.~A.Soward, Dr.~O.Podvigina, Prof.~U.Frisch,
Prof.~C.Jones and Dr.~A.Gilbert for numerous helpful discussions.
I would like to thank Prof.~K.Zhang for pointing me out
that boundary conditions \rf{Hboundary} are not always physically adequate,
and for informing me about the paper Roberts and Zhang (2000).
Most of the work on this project has been carried out during my visits
to the University of Exeter (UK) in May -- July 2002 and in January --
April 2004. I am grateful to the Royal Society for their support of the visits.
Some computations were performed using the facilities
provided by the program ``Simulations Interactives et Visualisation en
Astronomie et M\'ecanique (SIVAM)" at Observatoire de la C\^ote d'Azur,
France. My visits there were supported by the French Ministry of Education.
This work has been partly financed by the grant from the Russian
Foundation for Basic Research 04-05-64699.

\appendix
\renewcommand{\thesection}{\Alph{section}.}
\setcounter{equation}{0}
\renewcommand{\theequation}{\thesection\arabic{equation}}
\section{Asymptotic expansion of magnetic modes\hfill\break
and the associated eigenvalues for flows,\hfill\break
symmetric about the vertical axis}

Complete formal asymptotic expansions of magnetic eigenmodes and
associated eigenvalues are constructed here in the form \rf{Hseries},
\rf{lambdaseries} for small values of the scale ratio $\epsilon$.
Our basic assumption is that the problem is generic, i.e. the kernel
of the adjoint operator ${\cal L}^*$ is spanned by constant vector fields,
whose vertical component vanishes. The remaining assumptions concerning
the flow are listed in the first paragraph of Section~2.

Evidently, $\lad{\cal L}{\bf X}\rad=0$. Let us prove that the basic assumption is equivalent
to the assumption that $\cal L$, restricted to the subspace of vector fields
which have a zero horizontal mean and satisfy \rf{Hperiod} and \rf{Hboundary},
is invertible. An equation ${\cal L}{\bf X}=\bf F$ is solvable, if and only if
$\bf F$ is orthogonal to vector fields from ker\,${\cal L}^*$.
By the assumption, if $\lad{\bf F}\rad=0$, this condition is satisfied.
Let us show that $\bf X$ satisfying $\lad{\bf X}\rad=0$ can be found.
Suppose there exists
${\bf a}\in\ker{\cal L},\ \lad{\bf a}\rad=0$. ${\cal L}^*$ is an elliptic
operator; for the boundary conditions \rf{Hperiod} and \rf{Hboundary} it has
a point spectrum and its eigenvectors ${\bf a}^*_k$,
$${\cal L}^*{\bf a}^*_k=\xi^*_k{\bf a}^*_k;\quad\xi^*_k\ne0\hbox{ for }k\ge3,$$
constitute a complete basis (assuming for the sake of simplicity that it
does not have generalised eigenvectors). Expand $\bf a$ in this basis:
$${\bf a}=\sum_{k\ge3}c_k{\bf a}^*_k
=\sum_{k\ge3}{c_k\over\xi^*_k}{\cal L}^*{\bf a}^*_k.$$
Scalar multiplying this equation by $\bf a$ one finds $|{\bf a}|^2=0$.
(Alternatively, an argument similar to that of Jones and Roberts, 2000,
can be put forward: averaging a horizontal component of the equation
for a small-scale magnetic mode,
$${\cal L}{\bf a}_k=\xi_k{\bf a}_k,$$
for the assumed boundary conditions one finds $0=\xi_k\lad{\bf a}_k\rad$,
implying that all small-scale magnetic modes with a non-zero horizontal mean
belong to ker\,$\cal L$.)
Thus any non-zero vector field from the kernel of $\cal L$ has a non-zero
horizontal average. Consequently, it is possible to enforce
$\lad{\bf X}\rad=0$, subtracting from $\bf X$ the appropriate vector
from ker\,$\cal L$. This concludes the demonstration of equivalence
of the two assumptions.

Denote (in accordance with Section 2)
$$\H{n}=\lad\h{n}\rad;\quad\G{n}=\h{n}-\lad\h{n}\rad.$$
A mode is supposed to be $2\pi/L_j$-periodic in $x_j$ for $j=1,2$ and to
satisfy \rf{Hboundary} on the horizontal boundaries. Consequently, we assume
that each $\h{n}$ in \rf{Hseries} satisfies \rf{Hperiod} and \rf{Hboundary}.
Since $\lad\h{n}\rad$ automatically satisfies them, $\G{n}$ also does.

Substituting the series \rf{Hseries} and the expression for the spatial gradient
\rf{modnabla} into the solenoidality condition \rf{Hsolenoid} and expanding
the result, find
$$\sum_{n=0}^\infty\left(\nabla_{\bf y}\cdot\H{n-1}+
\nabla_{\bf x}\cdot\G{n}+\nabla_{\bf y}\cdot\G{n-1}\right)\epsilon^n=0$$
(by definition, $\H{n}=\G{n}\equiv0\ \forall n<0$). This defines
a hierarchy of equations. The mean of the equation at order $n+1$ is
$$\nabla_{\bf y}\cdot\H{n}=0;$$
the equation at order $n$ after subtraction of the mean becomes
$$\nabla_{\bf x}\cdot\G{n}+\nabla_{\bf y}\cdot\G{n-1}=0.$$

Substituting \rf{Hseries}-\rf{modnabla} into the eigenvalue
equation \rf{eigeneq}, obtain its series representation:
$$\left.\sum_{n=0}^\infty\right[{\cal L}\G{n}+\eta\left(
2(\nabla_{\bf x}\cdot\nabla_{\bf y})\G{n-1}
+\nabla^2_{\bf y}(\H{n-2}+\G{n-2})\right)$$
\begin{equation}
+\nabla_{\bf x}\times({\bf v}\times\H{n})
+\nabla_{\bf y}\times({\bf v}\times(\H{n-1}+\G{n-1}))
\label{hierarchy}\end{equation}
$$\left.-\sum_{m=0}^n\lmb{n-m}(\H{m}+\G{m})\right]\epsilon^n=0.$$
This yields a hierarchy of equations of the form
\begin{equation}
{\cal L}\G{n}={\bf F}^{(n)}({\bf x},{\bf y}).
\label{Leqn}\end{equation}
By the basic assumption this equation is solvable for $\G{n}$
if and only if \hbox{$\lad{\bf F}^{(n)}({\bf x},{\bf y})\rad=0$.}

Equations \rf{Leqn} arising from \rf{hierarchy} can be
solved successively in all orders by the following two-step procedure:\\
1. for a given $n$, from the solvability condition $\H{n-2}$ is determined and
the solvability condition is thus satisfied;\\
2. the resulting equation for this $n$ is solved for $\G{n}$.

\underline{$i$. $n=0$.} \rf{Leqn} reduces to
$${\cal L}\G{0}+(\H{0}\cdot\nabla_{\bf x}){\bf v}=
\lmb{0}(\H{0}+\G{0}).$$
The mean horizontal part of this equation is $0=\lmb{0}\H{0}$
implying $\lmb{0}=0$ (we consider modes which are not
predominantly small-scale, i.e. such that $\H{0}\ne0$).

By linearity, the solution with a zero horizontal mean
to the resulting equation admits a representation
\begin{equation}
\G{0}=\sum_{k=1}^2\Sk({\bf x})H_k^{(0)}({\bf y}),
\label{G0rep}\end{equation}
where $\Sk({\bf x})$ are solutions to the first auxiliary problem
\rf{Sproblem}, satisfying \rf{Hperiod} and \rf{Hboundary}.
The solvability condition for \rf{Sproblem} is verified,
since $\displaystyle\lad-\partial{\bf v}/\partial x_k\rad=0$ due to periodicity
of $\bf v$. In view of \rf{Skernel} equivalent to \rf{Sproblem},
$\Sk+{\bf e}_k$ span the kernel of $\cal L$.

Since $\bf v$ is symmetric about the vertical axis, the domain of $\cal L$ is
a direct sum of two proper subspaces, one of which is comprised of vector fields,
{\it symmetric about the vertical axis}:
$$H_1(-x_1,-x_2,x_3)=-H_1(x_1,x_2,x_3),$$
$$H_2(-x_1,-x_2,x_3)=-H_2(x_1,x_2,x_3),$$
$$H_3(-x_1,-x_2,x_3)=H_3(x_1,x_2,x_3);$$
and the second -- of the ones, {\it antisymmetric about the vertical axis}:
$$H_1(-x_1,-x_2,x_3)=H_1(x_1,x_2,x_3),$$
$$H_2(-x_1,-x_2,x_3)=H_2(x_1,x_2,x_3),$$
$$H_3(-x_1,-x_2,x_3)=-H_3(x_1,x_2,x_3).$$
Consequently, $\Sk$ is antisymmetric about the vertical axis.

Divergence of \rf{Sproblem},
\begin{equation}
\nabla^2(\nabla\cdot\Sk)=0,
\label{lapdivS}\end{equation}
suggests solenoidality of $\Sk$. Because of boundary conditions \rf{Hboundary}
for $\Sk$ and \rf{impenetr} for the flow, the vertical component of
\rf{Sproblem} implies $\left.\partial^2S^{(k)}_3/\partial^2x_3\right|_{x_3=0,\pi}=0$.
Consequently, $\nabla\cdot\Sk$ satisfies
$${\partial\over\partial x_3}\nabla\cdot\Sk=0$$
on horizontal boundaries. Evidently it has the same periodicity \rf{Hperiod}
as the flow. Multiplying \rf{lapdivS} by $\nabla\cdot\Sk$ and
integrating over the periodicity box, one obtains
$\la|\nabla(\nabla\cdot\Sk)|^2\ra=0\ \Rightarrow\ \nabla\cdot\Sk={\rm constant}$.
Integration of this equation over the periodicity box
demonstrates solenoidality of $\Sk$.

\underline{$ii$. $n=1$.} \rf{Leqn} takes now the form
$${\cal L}\G{1}+2\eta(\nabla_{\bf x}\cdot\nabla_{\bf y})\G{0}
+(\H{1}\cdot\nabla_{\bf x}){\bf v}$$
$$+\nabla_{\bf y}\times({\bf v}\times\G{0})
-({\bf v}\cdot\nabla_{\bf y})\H{0}=\lmb{1}(\H{0}+\G{0}).$$
The mean horizontal part of this equation is
$${\cal P}\nabla_{\bf y}\times\sum_{k=1}^2\la{\bf v}\times\Sk\ra
H_k^{(0)}=\lmb{1}\H{0}.$$
Since $\bf v$ is symmetric about the vertical axis and $\Sk$
is antisymmetric,
$$\la{\bf v}\times\Sk\ra_3=0\quad\Rightarrow\quad\lmb{1}=0.$$
Thus, although in the system under considerations the $\alpha$-effect tensor
does not vanish entirely, it does not affect generation of the principal
component of the mean field.

Consequently, after \rf{G0rep} is substituted, the equation becomes
$${\cal L}\G{1}=-(\H{1}\cdot\nabla_{\bf x}){\bf v}
-\sum_{k=1}^2\sum_{m=1}^2\left(2\eta{\partial\Sk\over\partial x_m}
+{\bf e}_m\times({\bf v}\times(\Sk+{\bf e}_k))\right)
{\partial H_k^{(0)}\over\partial y_m}.$$
By linearity, its solution with a zero horizontal mean admits a representation
\begin{equation}
\G{1}=\sum_{k=1}^2\Sk({\bf x})H_k^{(1)}({\bf y})+\sum_{k=1}^2\sum_{m=1}^2
\Gmk({\bf x}){\partial H_k^{(0)}\over\partial y_m}({\bf y}),
\label{G1rep}\end{equation}
where vector fields $\Gmk$ are solutions to the second auxiliary
problem \rf{Gproblem} satisfying the boundary conditions \rf{Hperiod} and \rf{Hboundary}.
Since $\bf v$ and $\Sk$ are symmetric and antisymmetric about the vertical
axis, respectively, and $\lad{\bf v}\rad=0$, the average of horizontal
components of the right-hand side of \rf{Gproblem} vanishes,
implying solvability of \rf{Gproblem}.

Due to the antisymmetry of $\Sk$ and the symmetry of $\bf v$, the
right-hand side of \rf{Gproblem} is symmetric about the vertical axis, and
since $\cal L$ preserves the symmetry and antisymmetry, $\Gmk$ are
symmetric about the vertical axis.

Let us show \rf{Gsolenoid}. Denote by $\bf\Phi$ a vector potential of $\bf v$:
$${\bf v}=\nabla\times{\bf\Phi}.$$
From \rf{Sproblem},
\begin{equation}
{\bf v}\times\Sk=\eta\nabla\times\Sk
-{\partial{\bf\Phi}\over\partial x_k}+\nabla\psi^{(k)}+{\bf C}^{(k)}.
\label{vcrossS}\end{equation}
Here $\bf\Phi$ and $\psi^{(k)}$ have the flow's periodicity \rf{vperiod}, and
${\bf C}^{(k)}$ is a constant vector. Now \rf{Gproblem} can be represented as
$${\cal L}\Gmk+\eta\left({\partial\Sk\over\partial x_m}
+\nabla\Smk\right)+\nabla\times((\Phi_k-\psi^{(k)}){\bf e}_m)
+{\bf e}_m\times{\bf C}^{(k)}=0.$$
Taking divergence of this equation, find
\begin{equation}
\nabla^2(\nabla\cdot\Gmk+\Smk)=0.
\label{lapdivG}\end{equation}
Because of boundary conditions \rf{Hboundary} for $\Sk$ and $\Gmk$, and
\rf{impenetr} for the flow, the vertical component of \rf{Gproblem} implies
$\left.\partial^2\Gamma^{(m,k)}_3/\partial^2x_3\right|_{x_3=0,\pi}=0$.
Consequently, the quantity $\phi=\nabla\cdot\Gmk+\Smk$ satisfies
$\partial\phi/\partial x_3=0$ on horizontal boundaries. Evidently,
it has the same periodicity \rf{Hperiod} as the flow. Multiplying \rf{lapdivG}
by $\phi$ and integrating over the periodicity box, obtain
$\la|\nabla\phi|^2\ra=0\ \Rightarrow\ \phi={\rm constant}$. Integration
of this equation over the periodicity box demonstrates \rf{Gsolenoid}.

\underline{$iii$. $n=2$.} \rf{Leqn} takes the form
$${\cal L}\G{2}+\eta\left(2(\nabla_{\bf x}\cdot\nabla_{\bf y})\G{1}
+\nabla^2_{\bf y}(\H{0}+\G{0})\right)$$
$$+(\H{2}\cdot\nabla_{\bf x}){\bf v}+\nabla_{\bf y}\times({\bf v}\times\G{1})
-({\bf v}\cdot\nabla_{\bf y})\H{1}=\lmb{2}(\H{0}+\G{0}).$$
The mean horizontal part of this equation is \rf{eddyeigen}.
Its solution \rf{Fourierharm}, \rf{toysolved} has been derived in Section 2.
Thus $\G{0}$ is determined together with $\H{0}$ from \rf{G0rep}.
The equation for $\G{2}$ can now be solved
by the general procedure outlined in $iv$ for $n\ge2$.

\underline{$iv$. $n=N>2$.} From equations for $n<N$ one finds:

\noindent
$\bullet$ vector fields $\G{n}$ for $n<N-2$, in particular
\begin{equation}
\G{n}({\bf x},{\bf y})={\bf g}^{(n)}({\bf x}){\rm e}^{{\rm i}\bf q\cdot y};
\label{Gfact}\end{equation}

\noindent
$\bullet$
\vspace{-\baselineskip}
\begin{equation}
\H{n}=0\quad\hbox{for }0<n<N-2;
\label{allzeros}\end{equation}

\noindent
$\bullet$ representations for $\G{n}$ of the form
\begin{equation}
\G{n}=\sum_{k=1}^2\Sk({\bf x})H_k^{(n)}({\bf y})+\sum_{k=1}^2\sum_{m=1}^2
\Gmk({\bf x}){\partial H_k^{(n-1)}\over\partial y_m}({\bf y})
+\Q{n}({\bf x}){\rm e}^{{\rm i}\bf q\cdot y}
\label{Gnrep}\end{equation}
for $n=N-1$ and $n=N-2$ with known vector fields $\Q{n}$, $\lad\Q{n}\rad=0$
(note \rf{G0rep} and \rf{G1rep} are particular cases of \rf{Gnrep}
with $\Q{0}=\Q{1}=0$);

\noindent
$\bullet$ $\lmb{n}$ for $n<N$.

The mean horizontal part of \rf{Leqn} is
\begin{equation}
({\cal M}-\lmb{2})\H{N-2}-\lmb{N}\H{0}
=-{\cal P}(\nabla_{\bf y}\times\la{\bf v}\times\Q{N-1}\ra)
+\sum_{m=1}^{N-3}\lmb{N-m}\H{m};
\label{lambdaN}\end{equation}
its right-hand side is known.
In view of \rf{Gfact}-\rf{Gnrep}, dependence of the right-hand side of
\rf{lambdaN} on the slow variable is via the factor ${\rm e}^{{\rm i}\bf q\cdot y}$.
Projecting this equation out in the direction
of $\H{0}$ in the space of bounded solenoidal vector fields of slow variables,
whose vertical component vanishes, one uniquely determines $\lmb{N}$.

In the complementary invariant subspace in the space of solenoidal
vector fields of slow variables only, whose vertical
component vanishes, the operator ${\cal M}-\lmb{2}$ is invertible.
Consequently, $\H{N-2}=0$ up to an arbitrary multiple of $\H{0}$,
which can be neglected (this is a normalisation condition).
Therefore, $\G{N-2}$ is now also determined.

The resulting equation for $\G{N}$ becomes
$${\cal L}\G{N}=-\eta\left(2(\nabla_{\bf x}\cdot\nabla_{\bf y})\Q{N-1}
+\nabla^2_{\bf y}\G{N-2}\right.$$
$$+\left.2\sum_{k=1}^2\sum_{m=1}^2\left(
{\partial\Sk\over\partial x_m}{\partial H_k^{(N-1)}\over\partial y_m}
+\sum_{l=1}^2{\partial\Gmk\over\partial x_l}
{\partial^2H_k^{(N-2)}\over\partial y_l\partial y_m}\right)\right)
-(\H{N}\cdot\nabla_{\bf x}){\bf v}+({\bf v}\cdot\nabla_{\bf y})\H{N-1}$$
$$-\nabla_{\bf y}\times\left(\sum_{k=1}^2({\bf v}\times\Sk)H_k^{(N-1)}
+\sum_{k=1}^2\sum_{m=1}^2\lb{\bf v}\times\Gmk\rb
{\partial H_k^{(N-2)}\over\partial y_m}+\lb{\bf v}\times\Q{N-1}\rb\right)$$
$$+\sum_{m=0}^{N-2}\lmb{N-m}\G{m},$$
where it is denoted $\lb\bf f\rb\equiv f-\lad f\rad$.
Only terms involving $\H{N}$ and derivatives of
$H_k^{(N-1)}$ have not yet been determined in it.
$\bf x$-dependent prefactors in front of these terms are
identical to those in front of $\H{1}$ and derivatives of $H^{(0)}_k$,
respectively, in the equations for $n=0$ and $n=1$. Therefore,
$\G{N}$ admits the required representation \rf{Gnrep}.
(An equation for $\Q{N}$ is obtained by replacing $\G{N}$ by $\Q{N}$ and
omitting all terms, involving $\H{N}$ and derivatives of $\H{N-1}$.)

Furthermore, it can be shown that $\Q{N}$ and $\G{N}$ are
symmetric about the vertical axis for odd $N$, and antisymmetric for even $N$.
Consequently, from \rf{lambdaN},
$${\cal P}(\nabla_{\bf y}\times\la{\bf v}\times\Q{N-1}\ra)=0
\quad\Rightarrow\quad\lmb{N}=0\hbox{ for odd }N.$$

Thus complete formal expansions of magnetic modes \rf{Hseries} and
the associated eigenvalues \rf{lambdaseries} have been constructed; they
satisfy \rf{Fourierharm}, \rf{toysolved}, \rf{allzeros} and \rf{Gfact}.

\section{Computation of elements\hfill\break
of the magnetic eddy diffusivity tensor}
\setcounter{equation}{0}

Here we discuss strategies for evaluation of elements of the MED tensor,
requiring to solve a smaller number of elliptic PDE's
than for the direct approach whereby all auxiliary problems are solved,
and present alternative expressions for elements of the MED tensor, where
integration required for averaging of the cross-products
of $\bf v$ and $\Gmk$ is partially performed.

As derived in Appendix A, elements of the MED tensor are equal to
$\la{\bf v}\times\Gmk\ra_l$. For flows symmetric about
the vertical axis in a layer, considered in this paper, $l=3$ and $m,k=1,2$. For
parity-invariant space-periodic flows, auxiliary problems are represented by the same
equations \rf{Sproblem} and \rf{Gproblem}, but now $l,m,k=1,2,3$.
To evaluate the MED tensor, the first and second auxiliary problems were solved
by Lanotte\al(2000), Zheligovsky\al(2001) and Zheligovsky and Podvigina (2003).
This approach requires solution of 6 elliptic PDE's ${\cal L}{\bf X}=\bf F$
for flows symmetric about the vertical axis in a layer, and 12 such PDE's
for parity-invariant space-periodic flows.

Let ${\bf W}^{(l)}$ be a solution to the auxiliary problem
\begin{equation}
{\cal L}^*{\bf W}^{(l)}={\bf v}\times{\bf e}_l.
\label{Lstareq}\end{equation}
The solvability condition for this equation is orthogonality of the right-hand
side to the kernel of $\cal L$.
For flows symmetric about the vertical axis in a layer, $l=3$ and
${\bf W}^{(l)}$ satisfies \rf{Hperiod} and \rf{Hboundary}. Solvability is implied
by the fact that $\bf v$ is symmetric about the axis, and ker\,$\cal L$
is spanned by $\Sk+{\bf e}_k$, which are antisymmetric about the axis.
For parity-invariant space-periodic flows, ${\bf W}^{(l)}$ is space-periodic;
solvability of \rf{Lstareq} follows from parity-invariance of $\bf v$ and
parity-antiinvariance of $\Sk+{\bf e}_k$ spanning ker\,$\cal L$.

Evidently, from \rf{Lstareq} and \rf{Gproblem}
$$\la{\bf v}\times\Gmk\ra_l=-\la{\cal L}^*{\bf W}^{(l)}\cdot\Gmk\ra
=-\la{\bf W}^{(l)}\cdot{\cal L}\Gmk\ra$$
$$=\left\la{\bf W}^{(l)}\cdot\left(2\eta{\partial\Sk\over\partial x_m}
+{\bf e}_m\times({\bf v}\times(\Sk+{\bf e}_k))\right)\right\ra.$$
Thus, it is enough to solve one auxiliary problem \rf{Lstareq} instead of four
problems \rf{Gproblem} for flows symmetric about the vertical axis in a layer,
and three auxiliary problems \rf{Lstareq} instead of nine problems \rf{Gproblem}
for parity-invariant space-periodic flows. Numerical complexity of problems
\rf{Lstareq} and \rf{Gproblem} is the same, since the operators $\cal L$
and ${\cal L}^*$ have the same spectrum. Thus, amount of computations can be
reduced twice by the use of the auxiliary vector fields ${\bf W}^{(l)}$.

In computations for parity-invariant space-periodic flows one can further
halve the number of problems to be solved. The curl of \rf{Lstareq} is
\begin{equation}
\eta\nabla^2{\bf R}^{(l)}-\nabla\times({\bf v}\times{\bf R}^{(l)})-
{\partial{\bf v}\over\partial x_l}=0,
\label{Zproblem}\end{equation}
where ${\bf R}^{(l)}=\nabla\times{\bf W}_l$. From \rf{Zproblem} and \rf{Sproblem},
\begin{equation}
\eta\nabla^2{\bf A}^{(l)}+\nabla\times({\bf v}\times{\bf B}^{(l)})=0,
\label{Aproblem}\end{equation}
\begin{equation}
\eta\nabla^2{\bf B}^{(l)}+\nabla\times({\bf v}\times{\bf A}^{(l)})
=-{\partial{\bf v}\over\partial x_l},
\label{Bproblem}\end{equation}
where it is denoted
\begin{equation}
{\bf A}^{(l)}={1\over2}({\bf S}^{(l)}+{\bf R}^{(l)}),\quad
{\bf B}^{(l)}={1\over2}({\bf S}^{(l)}-{\bf R}^{(l)}).
\label{ABnota}\end{equation}

Thus, it is in fact necessary to solve only one equation from the pair
\rf{Aproblem}, \rf{Bproblem}, e.g. \rf{Bproblem} in which ${\bf A}^{(l)}$
can be regarded as a notation for
$\eta^{-1}(-\nabla^2)^{-1}\nabla\times({\bf v}\times{\bf B}^{(l)})$,
since in the space of Fourier coefficients computation of the inverse
Laplacian is a simple operation. After ${\bf B}^{(l)}$ and thus ${\bf A}^{(l)}$
are determined, subsequently ${\bf S}^{(l)}$ and ${\bf R}^{(l)}$ can be found
from \rf{ABnota}, and from \rf{Lstareq}
$${\bf W}^{(l)}=(-\nabla^2)^{-1}\nabla\times{\bf R}^{(l)}+{1\over\eta}\nabla
(-\nabla^2)^{-2}\nabla\cdot\left({\bf v}\times({\bf R}^{(l)}+{\bf e}_l)\right).$$
However, computation of the left-hand side of \rf{Bproblem} in terms
of ${\bf B}^{(l)}$ requires now two inverse
and two direct FFT's, and thus merging the problems \rf{Lstareq} and
\rf{Sproblem} into a single equation \rf{Bproblem} with the notation
\rf{Aproblem} is not necessarily computationally advantageous.

If the flow possesses the ``translation antisymmetry":
$${\bf v}({\bf x})=-{\bf v}({\bf x}+{\bf a})$$
for a constant vector $\bf a$ (for the flow \rf{planforms}, \rf{potential}
this holds for ${\bf a}=\pi/L_2{\bf e}_2\,$, if $p$ is odd or $\alpha_2=0$),
solutions to the problem \rf{Sproblem}
can be obtained from solutions to \rf{Lstareq}:
$$\Sk({\bf x})=\nabla\times{\bf W}^{(k)}(\bf x+a).$$

Alternative representations of elements of the MED tensor are derived
in what follows using \rf{Sproblem} and \rf{Gproblem}.
For $j\ne0$ (and $1\le m,k\le 2$ for the problem in a layer) denote
$$\zeta^{(m,k,j)}={\rm e}^{{\rm i}jx_m}(\Sk+{\bf e}_k+{\rm i}j\Gmk)/j^2.$$
From \rf{Sproblem} and \rf{Gproblem}
\begin{equation}
{\cal L}\zeta^{(m,k,j)}={\rm e}^{{\rm i}jx_m}\left(\eta\left(\Sk+{\bf e}_k+{\rm i}j\Gmk
+2{\partial\Gmk\over\partial x_m}\right)+{\bf e}_m\times({\bf v}\times\Gmk)\right).
\label{zetaproblem}\end{equation}
Since the average of horizontal components of the left-hand side of
\rf{zetaproblem} vanishes, this implies
$${\cal P}\left\la{\rm e}^{{\rm i}jx_m}\left(\eta\left(\Sk+{\bf e}_k+{\partial\Gmk\over\partial x_m}\right)
+{\bf e}_m\times({\bf v}\times\Gmk)\right)\right\ra=0$$
for any $j\ne0$, and thus
\begin{equation}
{\cal P}\int\int\left(\eta\left(\Sk+{\bf e}_k+{\partial\Gmk\over\partial x_m}\right)
+{\bf e}_m\times({\bf v}\times\Gmk)\right){\rm d}x_{3-m}\,{\rm d}x_3\,={\bf C}^{(m,k)}A_m.
\label{Cmk}\end{equation}
Here $x_{3-m}$ and $x_3$ are Cartesian coordinates in directions orthogonal to
$x_m$, integration is performed over a section $x_m=$constant of the box of
periodicity of the flow, $A_m$ is the area of the section, and $C^{(m,k)}$
is a constant vector. Integrating \rf{Cmk} in $x_m$ and normalising
by the volume of the box of periodicity, one finds
$${\bf C}^{(m,k)}=\eta{\bf e}_k+{\cal P}\la{\bf e}_m\times({\bf v}\times\Gmk)\ra,$$
whereby from \rf{Cmk}
\begin{equation}
{\cal P}\la{\bf e}_m\times({\bf v}\times\Gmk)\ra={1\over A_m}
{\cal P}\int\int\left(\eta\left(\Sk+{\partial\Gmk\over\partial x_m}\right)
+{\bf e}_m\times({\bf v}\times\Gmk)\right){\rm d}x_{3-m}\,{\rm d}x_3.
\label{element1}\end{equation}
The right-hand side of \rf{element1} can be computed for any $x_m$.
The $m$-th component of the vector in the right-hand side of \rf{element1}
vanishes as a consequence of \rf{Gsolenoid}. Elements of the MED tensor
$\la{\bf v}\times\Gmk\ra_3$, required in the problem for a layer, can be
determined from the only remaining horizontal component of \rf{element1}.

For a parity-invariant flow, periodic in all three dimensions,
the average of the left-hand side of \rf{zetaproblem} vanishes, implying
by analogy with the case of a layer
\begin{equation}
\la{\bf e}_m\times({\bf v}\times\Gmk)\ra={1\over A_m}
\int\int\left(\eta\left(\Sk+{\partial\Gmk\over\partial x_m}\right)
+{\bf e}_m\times({\bf v}\times\Gmk)\right){\rm d}x_p\,{\rm d}x_q.
\label{element1p}\end{equation}
Here $x_p$ and $x_q$ are Cartesian coordinates in directions orthogonal
to $x_m$, integration is performed over the section $x_m=$constant of
the box of periodicity of the flow.
The right-hand side of \rf{element1p} can be computed for any $x_m$.
The $m$-th component of the vector in the right-hand side of \rf{element1p}
vanishes as a consequence of \rf{Gsolenoid}. The $p$-th and $q$-th components
of $\la{\bf v}\times\Gmk\ra$, required in the three-dimensional space-periodic
problem, can be easily determined from this equation.

Another expression for $\la{\bf v}\times\Gmk\ra$, including the $m$-th
component, can be obtained from \rf{Gproblem} noticing, that vector fields
$\epsilon_{pql}x_q{\bf e}_p$ (no summation over repeating indices is assumed!)
solve \rf{Lstareq} (this observation is, of course, of no consequence
for determination of ${\bf W}^{(l)}$, since this vector field does not satisfy
the required periodicity condition \rf{Hperiod}). Here $\epsilon_{pql}$
is the permutation symbol:
$${\bf e}_p\times{\bf e}_q=\epsilon_{pql}{\bf e}_l,$$
the three indices $p,q$ and $l$ being distinct.
Multiplying \rf{Gproblem} by $x_q$, integrating over the box of periodicity
and normalising by the volume of the box, find
$$\la{\bf e}_q\times({\bf v}\times\Gmk)\ra={1\over A_q}
\int\int\left(\eta\left(2\delta^q_m\Sk+{\partial\Gmk\over\partial x_q}
\right)+{\bf e}_q\times({\bf v}\times\Gmk)\right){\rm d}x_l\,{\rm d}x_p$$
\begin{equation}
+\la x_q{\bf e}_m\times({\bf v}\times(\Sk+{\bf e}_k))\ra.
\label{element2}\end{equation}
Here $\delta^q_m$ is the Kronecker symbol and the integral over a section
of the box of periodicity of the flow in the left-hand side can be evaluated
for any $x_m$. For $q=m$ \rf{element2} does not immediately reduce to
\rf{element1p}. To perform the reduction, note that in the three-dimensional
space-periodic problem ${\bf C}^{(k)}=0$, cross-premultiply \rf{vcrossS} by
$x_m{\bf e}_m$, integrate over the box of periodicity, normalise the result
by the volume of the box and subtract it from \rf{element2}.

Evidently, \rf{element2} also holds for flows in a layer
for $1\le p,q\le 2$ and $l=3$.

\pagebreak

\end{document}